\documentclass[epj]{svjour}
\usepackage{graphics}
\def\prn#1{{\left(#1\right)}}

\def\sbrk#1{{\left[#1\right]}}

\def\sbrk#1{{\left[#1\right]}}

\def\prn#1{{\left(#1\right)}}

\newcommand{\TDOne}{${\rm 5d6s \:} ^3{\rm D}_1\:$}
\newcommand{\SPOne}{${\rm 6s6p \:} ^1{\rm P}_1\:$}
\newcommand{\SSZeroToTDOne}{\mbox{${\rm 6s^2}~ ^1{\rm S}_0~\rightarrow ~{\rm 5d6s
}~ ^3{\rm D}_1\:$}}

\def\clebsch#1#2#3#4#5#6{\langle #1 \, #2 \, #3 \, #4 \vert #5 \, #6 \rangle}

\begin{document}
\title{Towards measuring nuclear-spin-dependent and isotopic-chain atomic parity violation in Ytterbium}
\author{K. Tsigutkin\inst{1} \and J. E. Stalnaker\inst{1}\thanks{\emph{Present address:} National
Institute of Standards and Technology, 325 S. Broadway Boulder, CO
80305-3322}%
\and D. Budker\inst{1,2} \and S. J. Freedman\inst{1,2} \and V. V.
Yashchuk\inst{3}
}                     
%
%
\institute{Department of Physics, University of California,
Berkeley, CA 94720-7300 \and Nuclear Science Division, Lawrence
Berkeley National Laboratory, Berkeley CA 94720 \and Advanced Light
Source Division, Lawrence Berkeley National Laboratory, Berkeley CA
94720}
\date{Received: date / Revised version: date}
%
\abstract{We discuss experiments aimed at measurements of atomic
parity nonconservation (PNC) effects in the $\SSZeroToTDOne$
transition (408 nm) in atomic Ytterbium (Z=70). According to
theoretical predictions, the PNC-induced E1 amplitude of this
transition is $\sim 100$ times larger than the analogous amplitude
in Cs. Such an experiment will determine differences in PNC effects
between different hyperfine components for odd-neutron-number Yb
isotopes and, thereby, will allow measurements of the nuclear
anapole moments in nuclei with unpaired neutrons. In addition,
measurements of PNC in different isotopes would give information on
neutron distributions within the nuclei. The apparatus designed and
built for this experiment is described, and results of measurements
towards understanding of systematic effects influencing the
accuracy, and the current status of the ongoing PNC measurements are
presented.
} 
\authorrunning{K. Tsigutkin \textit{et al}}
\titlerunning{Towards measuring nuclear-spin-dependent and isotopic-chain atomic parity violation...}
\maketitle
\section{Introduction}
\label{Sect:intro} 
\begin{figure}
\resizebox{0.45\textwidth}{!}{%
  \includegraphics{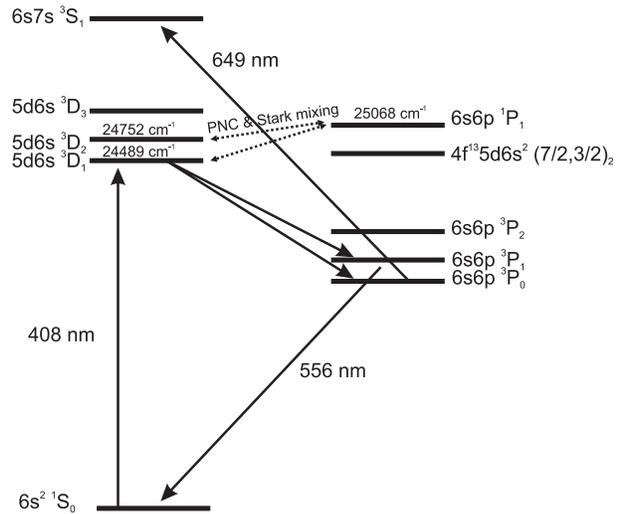}
}
\caption{Low-lying energy eigenstates of Yb and transitions relevant
to the PNC experiment.}
\label{fig energy}       
\end{figure}

Atomic ytterbium (Yb) was proposed as a system
for measuring the effects of parity nonconservation (PNC) by D.
DeMille \cite{Dem95} who pointed out a rather non-obvious
enhancement of the PNC transition amplitude arising due to
configuration mixing in addition to relative proximity of the
excited \TDOne and \SPOne states of opposite nominal parity (Fig.
\ref{fig energy}). The PNC-induced \SSZeroToTDOne\ transition
amplitude has been calculated \cite{Dem95,Por95,Das97} to be two
orders of magnitude larger than the PNC amplitude in cesium where
high-precision measurements have been carried out
\cite{Woo99,Gue2005}.

While the accuracy of the theoretical calculations of PNC effects in
Yb is at a $\sim 20 \%$ level \cite{Por95,Das97}, there is little
hope that it will ever reach sub-one-percent levels achieved for Cs
and other single-electron systems (see, for example, review
\cite{Gin2004}). Therefore, the motivation for PNC experiments in Yb
is not testing the Standard Model via a comparison of high-precision
theory and experiment, but rather probing low-energy nuclear physics
by comparing PNC effects on either a chain of naturally occurring Yb
isotopes ($^{168}$Yb-0.13\%, $^{170}$Yb-3.04\%, $^{171}$Yb-14.28\%,
$^{173}$Yb-16.13\%, $^{174}$Yb-31.83\%, $^{176}$Yb-12.76\%), or in
different hyperfine components of the same odd-neutron-number
isotope ($^{171}$Yb-nuclear spin $I=1/2$ or $^{173}$Yb-$I=5/2$). The
former would yield information on the mean-square neutron radii in
the chain of isotopes, while the latter would probe
nuclear-spin-dependent PNC, which is sensitive to the nuclear
\emph{anapole moments} (see, for example, reviews
\cite{Gin2004,Hax2001}) that arise from weak interactions between
the nucleons. Both of these types of measurements probe nuclear
physics that is exceedingly difficult to access by other means, and
neither relies on high-precision theory.

\section{Preliminary experiments}
\label{Sect:Prelim_Expts} Experimental work towards an Yb-PNC
measurement has started at Berkeley several years ago, and, in its
initial phase, was directed towards measurement of various
spectroscopic properties of the Yb system of direct relevance to the
PNC measurement, including determination of radiative lifetimes
\cite{Bow96}, measurement of the Stark-induced amplitudes, hyperfine
structure, isotope shifts, and dc-Stark shifts of the \mbox{${\rm
6s^2}~ ^1{\rm S}_0~\rightarrow ~{\rm 5d6s }~ ^3{\rm D}_{1,2}\:$}
transitions, and the quadrupole moment of the latter transition
($\lambda= 404\ $nm) \cite{Bow99}, and the forbidden magnetic-dipole
(M1) amplitude of the former transition ($\lambda= 408\ $nm)
\cite{Sta2002}. For the measurement of the forbidden magnetic-dipole
transition amplitude ($\approx 10^{-4}\ \mu_B$), we used the
M1-(Stark-induced)E1 interference technique. A simple atomic system
where transition amplitudes and interferences are well understood
has proven useful for gaining insight into curious Jones dichroism
effects that had been studied in condensed-matter systems at extreme
conditions and whose origin had been a matter of debate (see Ref.
\cite{Bud2003Jones} and references therein).

Most recently \cite{Sta2006}, we reported on an experimental and
theoretical study of the dynamic (ac) Stark effect on the
\SSZeroToTDOne forbidden transition. A general framework for
parameterizing and describing off-resonant ac-Stark shifts was
presented. A model was developed to calculate spectral line shapes
resulting from resonant excitation of atoms in an intense standing
light-wave in the presence of off-resonant ac-Stark shifts. A
bi-product of this work was an independent determination (from the
saturation behavior of the 408-nm transition) of the Stark
transition polarizability, which was found to be in agreement with
our earlier measurement \cite{Bow99}.

The present incarnation of the Yb PNC experiment (described below)
involves a measurement using an atomic beam. An alternative approach
would involve working with a heat-pipe-like vapor cell. Various
aspect of such an experiment were investigated, including
measurements of collisional perturbations of relevant Yb states
\cite{Kim99}, nonlinear optical processes in a dense Yb vapor with
pulsed UV-laser excitation \cite{DeB2001}, and an altogether
different scheme for measuring PNC via optical rotation on a
transition between excited states \cite{Kim2001}.

\section{Experimental technique for the PNC measurement}
\label{Sect:Expt_Technique} The general idea of the present Yb-PNC
experiment is to excite the forbidden 408-nm transition (Fig.
\ref{fig energy}) with resonant laser light in the presence of a
static electric field. The purpose of the electric field is to
provide a reference Stark-induced transition amplitude, much larger
than the PNC amplitude. In such a Stark-PNC interference method
\cite{Bou75,Con79}, one is measuring the part of the transition
probability that is linear in both the reference Stark-induced
amplitude and the PNC amplitude. In addition to enhancing the
PNC-dependent signal, employing the Stark-PNC interference technique
provides for all-important reversals allowing one to separate the
PNC effects from various systematics.

\begin{figure}
\resizebox{0.45\textwidth}{!}{%
  \includegraphics{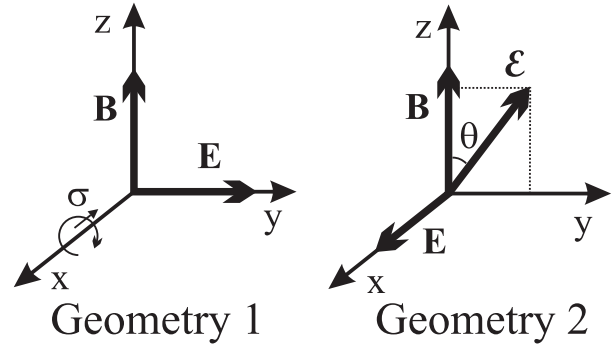}
} \caption{Possible orientation of fields for PNC-Stark interference
experiment. In both cases, light is applied collinearly with x.}
\label{fig PNCGeoms}
\end{figure}

We begin the discussion of the experiment by considering the
rotational invariant to which the PNC-Stark-interference term is
proportional.  Since the effect violates parity and conserves
time-reversal invariance, we require the rotational invariant to be
P-odd and T-even.  The two configurations which have been used for
Stark-interference experiments of this type are shown in Fig.
\ref{fig PNCGeoms}.

Geometry One was used in the Cs Stark-interference experiment
\cite{Woo97} and Geometry Two was used in an early
Stark-interference experiment in Tl \cite{Dre85}.  The rotational
invariant for Geometry One is given by
\begin{equation}
\bf{\sigma} \cdot \prn{\bf{E} \times \bf{B}},\label{Eq:GeometryOne}
\end{equation}
where $\bf{\sigma}$ is the helicity of the excitation light. For
Geometry Two, the invariant is
\begin{equation}
\prn{\mathcal{E} \cdot
\bf{B}}\prn{\sbrk{\bf{E}\times\mathcal{E}}\cdot
\bf{B}},\label{Eq:GeometryTwo}
\end{equation}
where $\mathcal{E}$ is the electric field of the (linearly
polarized) light.

The two geometries have very different characteristics and are
sensitive to different sets of systematics.  Neither geometry is
clearly superior. One advantage of Geometry Two is an additional
suppression of spurious M1-Stark interference, one of the most
important systematic effects in precision PNC-Stark interference
experiments. In both Geometries One and Two, this effect can be
minimized by the use of a power-build-up cavity to generate a
standing light wave. Since a standing wave has no net direction of
propagation, $\bf{k}$, any transition rate which is linear in the M1
amplitude (proportional to $\bf{k}\times \mathcal{E}$) will cancel
out. However, this is the only source of suppression of the M1-Stark
interference in Geometry One. This can be seen by replacing
$\bf{\sigma}$ in the rotational invariant (\ref{Eq:GeometryOne})
with $\bf{k}$. This is a T-even, P-even rotational invariant which
arises from M1-Stark interference. Geometry Two, on the other hand,
is such that the M1 and Stark amplitudes are out of phase. Thus, the
M1 and Stark amplitudes do not interfere and therefore do not
produce additional interference terms which can change with the
changing fields and mimic the PNC-Stark interference. This geometry
can even be used in the absence of a power-build-up cavity (and in
fact was in the work of Ref. \cite{Dre85}).  Thus, the M1-Stark
interference term requires a misalignment of the fields along with a
net direction of propagation resulting from an imperfect standing
wave.  This additional reduction of M1-Stark interference has led us
to choose Geometry Two for the PNC-Stark experiment.

The rotational invariant (\ref{Eq:GeometryTwo}) highlights some of
the salient features of our experiment: the PNC effect reverses with
dc electric field ($\bf{E}$), and does not reverse with the magnetic
field ($\bf{B}$); however, the presence of the magnetic field is
necessary. In fact, the magnetic field should be strong enough so
that the transitions to different Zeeman components of the upper
state are resolved (because the PNC effect averaged over all Zeeman
components is zero). Another reversal is flipping the axis of the
light polarization with respect to the E-B plane, which also changes
the sign of the PNC effect.

\section{PNC signatures for even and odd isotopes}
\label{Sec:PNC_Signatures} Here we address briefly the effect of the
weak mixing between $^3 D_1$ and $^1 P_1$ states on the
\SSZeroToTDOne\ transition. In the present experiment, the forbidden
transition rate is enhanced by applying a DC electric field to Yb
atoms which mixes the $^3 D_1$ and $^1 P_1$ states. In order to
discriminate the weak mixing contribution to the transition rate, a
magnetic field is applied leading to Zeeman splitting of the
transition into several components. In the case of the Geometry 2
(Fig. \ref{fig PNCGeoms}), for the even ($I=0$) Yb isotopes, the
transition is split onto three components. For the odd isotopes, due
to the hyperfine splitting of the $^3 D_1$ state, the magnetic field
produces different Zeeman patterns for upper levels with different
values of the total angular momentum $F'$, but in each case, there
is an even number of Zeeman components.

A Stark-induced transition amplitude is generally expressed in terms
of real scalar ($\alpha$), vector ($\beta$), and tensor ($\gamma$)
transition polarizabilities \cite{Bou75,Bow99}, however, for the
case of a $J=0\rightarrow J'=1$ transition, only the vector
transition polarizability contributes. The transition amplitude
between states with total angular momenta and projections $F,M$ and
$F',M'$ takes the form\footnote{The $(-1)^q$ factor was
inadvertently omitted in some of our earlier publications. This has
not affected any of the quoted results.}
\begin{eqnarray}  \label{StAmp_general}
&&A^{Stark}_{FMF'M'}\\&&=i\beta_{FF'}(-1)^{q}\left(\bf{E}\times\vec{\mathcal{E}}\right)_{q}\langle
F,M,1,-q|F',M'\rangle~,\nonumber
\end{eqnarray}
where $q=M-M'$ labels the spherical component and
$\clebsch{F,}{M,}{1,}{-q}{F',}{M'}$ is a Clebsch-Gordan coefficient.
The transition amplitude arising due to PNC can be expressed as
\begin{eqnarray}  \label{PNC_Amp}
A^{PNC}_{FMF'M'}=i\xi_{FF'}(-1)^{q}\vec{\mathcal{E}}_{q}\langle
F,M,1,-q|F',M'\rangle~.
\end{eqnarray}
Here $\xi_{FF'}$ characterizes the PNC-induced electric-dipole
transition amplitude.

Assuming that the magnetic field is perpendicular to the electric
field and is strong enough to resolve the Zeeman components of the
transition, that the light propagates parallel to the electric
field, and selecting the quantization axis along the magnetic field,
we obtain the following results for the transition rates.
For isotopes with zero nuclear spin, the transition rates to Zeeman
sublevels $M'=0,\pm 1$ are:
\begin{eqnarray}  \label{even_rates}
\mathcal{R}_0=\frac{8\pi}{c}\mathcal{I}\sbrk{\beta^2
E^2\sin^2\theta+2\xi\,\beta E  \sin \theta \cos \theta}~,
\\\mathcal{R}_{\pm 1}=\frac{4\pi}{c}\mathcal{I}\sbrk{\beta^2
E^2\cos^2\theta-2\xi\,\beta E  \sin \theta \cos
\theta}~,\label{even_rates2}
\end{eqnarray}
where $\mathcal{I}$ is the light intensity. In these expressions, we
neglect the term in the transition rate which is quadratic in PNC
mixing.
\begin{figure}
\resizebox{0.45\textwidth}{!}{%
  \includegraphics{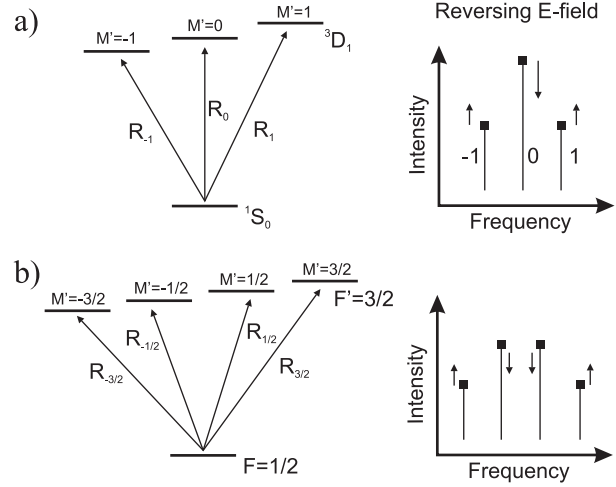}
} \caption{(a) Level diagram and Zeeman-component spectrum for $^1
S_0\rightarrow ^3D_1$ transition for even Yb isotopes. (b) Same for
$^1 S_0~F=1/2\rightarrow ^3D_1~F=3/2$ for $^{171}$Yb. The small
arrows indicate the effect of the Stark-PNC interference term -- a
small relative enhancement of some components and a suppression of
others -- that reverses with $\bf{E}$ and $\theta$.}
\label{Fig:lshape}
\end{figure}
For an arbitrary angle of the light polarization $\theta$, there are
generally three components of the transition as shown in Fig.
\ref{Fig:lshape}a. Neglecting Stark shifts, the distance between
adjacent peaks corresponds to an energy difference
\begin{equation}
\Delta E \approx g \mu_0 M' B\ ,
\end{equation}
where $g$ is the Land\'e factor, $\mu_0$ is the Bohr magneton, and
$B$ is the magnetic field. The effect of the Stark-PNC interference
is to change the relative strength of the two extreme components of
the transition with respect to the central component. Expressions
(\ref{even_rates},\ref{even_rates2}) show explicitly that the PNC
effect reverses with both $\vec{\bf{E}}$ and $\theta$.

As an example of a transition for an odd isotope, consider
$F=1/2\rightarrow F'=3/2$ of $^{171}$Yb (Fig. \ref{Fig:lshape}b).
Since the magnetic moment of the ground $^1S_0~F=1/2$ state is
solely due to the nucleus, the Zeeman splitting of this state is
three orders of magnitude smaller than that of the $^3D_1~F=3/2$
state and is unresolved at the magnetic-field strength of interest.
Therefore, the total rates for the transitions to the upper-state
Zeeman components $M'$ are sums of the respective rates. The vector
transition polarizability $\beta_{FF'}$ is related to that for an
isotope with zero nuclear spin ($\beta$) according to
\begin{equation}
\beta_{FF'}=\beta(-1)^{I+F}\sqrt{(2F'+1)(2F+1)}\left\{
\begin{array}{ccc}
1 & F' & I \\
F & 0 & 1 \\
\end{array} \right\}~,
\end{equation}
where the brackets denote a $6j$-symbol \cite{Sob92}. The distance
between peaks is determined by
\begin{equation}
\Delta E \approx g_{F'} \mu_0 M' B~,
\end{equation}
where $g_{F'}$ is given by
\begin{equation}
g_{F'}=g\cdot\frac{F'(F'+1)+J'(J'+1)-I(I+1)}{2F'(F'+1)}~.
\end{equation}
In calculating the PNC amplitude, one needs to take into account
both the nuclear-spin-independent effect (as in the case of the even
isotopes), as well as the nuclear spin-dependent contribution,
including the effect of the nuclear anapole moment. The PNC
signature for this transition is an asymmetry between the two pairs
of Zeeman components of the transition (Fig. \ref{Fig:lshape}b)
which, once again, reverses with $\vec{\bf{E}}$ and $\theta$.

In addition, we note that the $F=1/2\rightarrow F'=1/2$ transition
of $^{171}$Yb is not influenced by the PNC effect under the
discussed conditions. This transition is split by the magnetic field
into two Zeeman components, whose intensities must be invariant
under the B-field reversal. Thus, the PNC asymmetry cancels. The
total rates for these two Zeeman components turn out to be equal and
independent of the polarization angle $\theta$. Observation of this
transition gives us an additional handle to study systematics.

Using the theoretical value of $\xi\simeq 10^{-9}~ea_0$
\cite{Por95,Das97} combined with the measured $\beta\simeq 2\times
10^{-8}~ea_0/$(V/cm) \cite{Bow99,Sta2006}, we can estimate the
relative asymmetry in the line shapes. For example, for
$\mathcal{R}_0$ and $\theta=\pi/4$, and for $E=1\ $kV/cm, the
asymmetry is $\simeq 10^{-4}$, with the nuclear spin-dependent and
isotope-dependent effects expected at a few percent of that
(depending on a specific model).

\section{Apparatus and signals}
\label{Sec:Apparatus}
\begin{figure}
\resizebox{0.5\textwidth}{!}{%
  \includegraphics{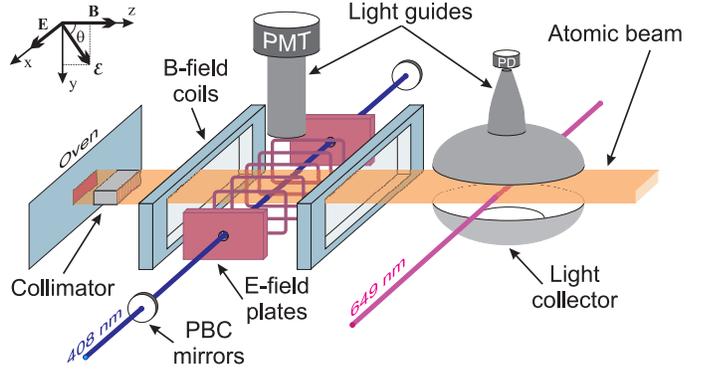}
} \caption{Schematic of the present Yb-PNC apparatus. Not shown are
the atomic-beam oven and the vacuum chamber containing all the
depicted elements, except the photomultiplier (PMT), photodiode
(PD), and the photodiode light-guide. PBC--power buildup cavity.}
\label{fig PNC_apparatus}
\end{figure}
\begin{figure}
\resizebox{0.5\textwidth}{!}{%
  \includegraphics{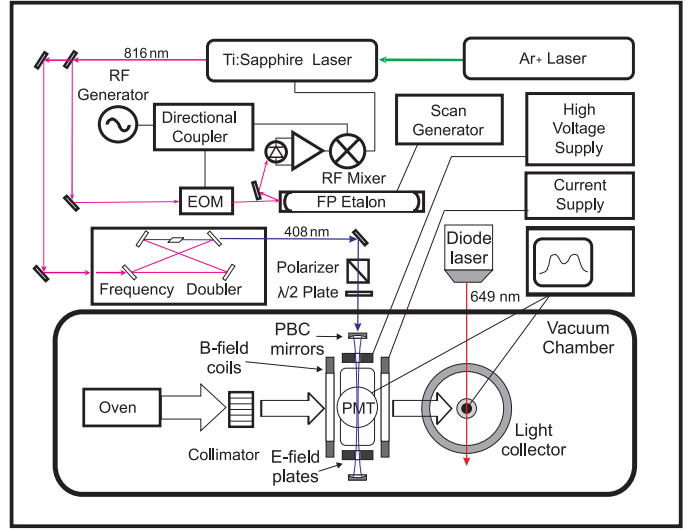}
} \caption{Schematic of the optical setup and the frequency-locking
systems. EOM--electro-optical modulator, FP-Fabry-Perot. The ``diode
laser'' producing 649-nm light consists of a commercial master laser
(New Focus Vortex) and a home-made injection-locked diode
amplifier.} \label{fig PNC_apparatus_optical}
\end{figure}

A schematic of the Yb-PNC apparatus is shown in Fig. \ref{fig
PNC_apparatus}. A beam of Yb atoms is produced (inside of a vacuum
chamber with a residual pressure of $\approx 5 \times 10^{-6}~{\rm
Torr}$) with an effusive source (not shown), which is a
stainless-steel oven loaded with Yb metal. The oven is outfitted
with a multi-slit nozzle providing initial collimation of the atomic
beam. The oven is operated with the front $\approx 100^{\circ}{\rm
C}$ hotter than the rear to avoid clogging of the nozzle.  The
typical operating temperatures are $ 500^{\circ}{\rm C}$ in the rear
and $600^{\circ}{\rm C}$ in the front. In order to reduce the
Doppler width of the 408-nm transition and the overlap between its
various hyperfine and isotopic components, a downstream external
vane collimator was installed. The collimator provides a collimation
angle of $\approx 1 ^\circ $ in the horizontal direction. This
reduces the Doppler width of the 408-nm transition to $\approx 12 \,
{\rm MHz}$. The transparency of the collimator is estimated to be
$\approx 90 \%$. The vane collimator is mounted on a movable
platform, allowing precise alignment of its angle relative to the
atomic beam during the experiment, as well as its removal from the
beam.

Downstream from the collimator, the atoms enter the main interaction
region where the Stark- and PNC-induced transition takes place. The
light at the transition wavelength of $408.345 \, {\rm nm}$ is
produced by doubling the output of a Ti:Sapphire laser (see
schematic of the optical arrangement, Fig. \ref{fig
PNC_apparatus_optical}). A Coherent $899$ Ti:Saphire laser is pumped
by $\approx 12 \,{\rm W}$ from a Spectra Physics 2080 argon-ion
laser operating on all lines. The Ti:Sapphire laser produces
$\approx 1.2 \, {\rm W}$ of light at $816 \, {\rm nm}$. This light
is frequency doubled using a commercial bow-tie resonator with a
Lithium-Triborate (LBO) crystal (Laser Analytical Systems Wavetrain
cw). The output of the frequency doubler is $\approx 80 \, {\rm
mW}$.

The 408-nm light is coupled into a power buildup cavity (PBC). We
operate the cavity in the confocal regime (mirror radii of curvature
and the mirror separation are 25 cm) in order to maintain a
relatively large diameter of the light beam by simultaneously
exciting many degenerate transverse modes. This is important for our
ability to control the ac-Stark shifts without compromising the size
of the signals. Precision optical mounts are used for PBC mirrors
with micrometer adjustments for the horizontal and vertical angles
and the pivot point of the mirror face. The mirrors are attached
with RTV adhesive to thin aluminum rings.  It was hoped that the
thin wall aluminum frame and RTV adhesive would reduce
birefringent-induced stress resulting from the adhesion as well as
from thermal expansion and contraction. The aluminum frame of one of
the mirrors is attached to a piezo-electric ceramic which is
attached to the optical mount. The three micrometer adjustments of
one of the optical mounts are effected with precision, vacuum
compatible pico-motors (New Focus 8302-v) to allow for adjustment of
the cavity alignment while in vacuum. The finesse of the PBC was
measured using the cavity-ring-down method to be $\mathcal{F} =
4240(70)$. The fraction of the incident light power that could be
coupled through the cavity was typically 10-18\%, limited to a large
extent by the losses in the mirrors (measured at about 240 ppm per
bounce). We note that the losses for the mirrors presently used in
this experiment are $\approx 40$ times larger than those for the
mirrors used in some other PBCs (e.g., those in Refs.\
\cite{Woo97,Hoo2001}). A reason for such relatively high losses is
that our wavelength is significantly shorter. (The losses are
expected to be significantly lower in the state-of-the-art mirrors
presently on order).

The Coherent 899 Ti:Sapphire laser is rated to have a short-term
instability (effective line width) of $500$~kHz.  In order to reduce
the line width of the laser (which is necessary for efficient
coupling into the PBC) an electro-optic modulator (EOM) was placed
inside the laser cavity. Changing the voltage across the EOM changes
the index of refraction of the crystal and therefore the
optical-path length. The EOM is capable of extremely fast response
with a $\approx 5$ MHz bandwidth, limited by the speed of the
electronics. The EOM is a double-crystal assembly (LINOS Optics PM
25 IR); the crystals are cut at Brewster's angle to minimize loss
and are compensated to prevent walk off of the beam. It is possible
to insert the EOM into the cavity and achieve lasing with minimal
realignment of the laser. The output of the laser drops $\approx
5-10\%$ when the EOM is added. The laser is locked to the PBC using
the fm sideband technique \cite{Drev83}.

We found that using the PBC as the ``master'' cavity leads to
frequency oscillations at acoustic frequencies. In order to remove
these oscillations, the resonant frequency of the cavity is locked
to a more stable confocal Fabry-Perot etalon, once again using the
fm sideband locking technique.  Thus, the stable (scannable) cavity
provides the master frequency, with the power-build-up cavity
serving as the secondary master for the laser.

The relative frequency of the laser light is determined using a
homemade spherical-mirror Fabry-Perot interferometer operating at
408 nm (not shown in Fig. \ref{fig PNC_apparatus_optical}). Since
the experiment relies on a detailed understanding of the spectral
line shape of the transition over a region of $\approx 100\ $MHz, a
frequency reference with closely spaced frequency markers is needed.
To this end, the interferometer is operated with the mirror spacing
chosen so that the transverse cavity modes overlap at frequency
intervals of $\Delta\nu_{res}=c/(2\, N \, L),$ with $N=7$ ($c$ is
the speed of light, and $L$ is the mirror separation)
\cite{Bud2000FP}. This allows one to achieve relatively closely
spaced frequency markers without making an excessively long cavity.
The interferometer consists of two mirrors each of which has a
radius of curvature of $R=50 \,{\rm cm}$ and the mirrors are
separated a distance 38.9 cm.  This results in a spacing between the
cavity resonances of 55.12 MHz. The cavity is made of invar and
placed in an evacuated, passively thermally stabilized enclosure.
The frequency drifts of the cavity resonances are $\approx 1$
MHz/min.

For the PNC measurements, we plan to take most of the data with an
electric-field magnitude in the interaction region of $1.5\ $kV/cm.
The electric field is generated with two copper electrodes with
dimensions 4.4x2.5x0.4$\ $cm$^3$ separated by a 6.8-cm gap. In the
inter-electrode gap, nine equidistant copper frames are placed for
producing a homogeneous voltage drop along the optical axis of the
cavity. These frames are connected between each other and the
electrodes through 10-M$\Omega$ 0.1\%-accuracy high-voltage
resistors forming a voltage divider. A 10-kV voltage is supplied to
the E-field plates by a SPELLMAN CZE1000R power supply modified to
allow computer controlled switching of the voltage polarity. During
preliminary diagnostic experiments where higher electric field was
desirable (see, for example, the data in Figs. \ref{fig signal_no_B}
and \ref{fig Signal_B}), the inter-electrode frames were removed and
the gap was reduced to 1 cm. Fields of $12\ $kV/cm can be reliably
obtained without breakdown.

The magnetic field is generated by a pair of rectangular coils
having dimensions: 12(width)x3.2(height)x1(depth) cm$^3$\ and
separated by a 1.4-cm gap. The coils are designed to produce a
uniform (1\% non-uniformity) axial magnetic field up to $100\ $G
within the region where the atomic beam intersects the waist of the
power-buildup cavity (see Fig. \ref{fig PNC_apparatus}). The coils a
powered by a computer-controlled power supply allowing modulation
and reversal of the magnetic field.

Light emitted from the interaction region at 556$\ $nm (Fig.
{\ref{fig energy}) resulting from the second step of the
fluorescence cascade following excitation to the \TDOne state is
collected with a light guide and detected with a photomultiplier
tube. Because of the field requirements in the interaction region,
it is difficult to achieve efficient light detection. In order to
improve the detection efficiency for the PNC experiment a separate
downstream detection region, dedicated to monitoring the number of
atoms making a transition is used. We utilize the fact that $65 \%$
of the atoms excited to the \TDOne state decay to the meta-stable
${\rm 6s6p} \: ^3{\rm P}_0$ state (Fig. \ref{fig energy}).  These
atoms are probed by resonantly exciting them (with 649-nm light) out
of the ${\rm 6s6p} \: ^3{\rm P}_0$ state to the ${\rm 6s7s} \:^3{\rm
S}_1$ state. The subsequent fluorescence is detected in a region
free of other experimental components.

The 649-nm light is derived from a single-frequency diode laser (New
Focus Vortex) producing $\approx 1.2\ $mW of cw output that is
amplified in home-made injection-locking system to $\approx 7\ $mW
of light sent into the detection region.

The detection region consists of two optically polished curved
aluminum reflectors. One reflector, which has hemispherical shape
with a radius of curvature of 8.9 cm, covers the upper hemisphere
with respect to the intersection of the 649-nm laser beam with the
atomic beam. The second reflector is parabolic and is located below
the beam intersection. This reflector is positioned in such a way
that the focal point of the parabola coincides with the center of
curvature of the upper reflector and with the beam intersection.
Fluorescent light from the atoms excited with the 649-nm light is
collected onto a plastic light guide through a 5-cm diameter opening
in the upper reflector. The light-guide is a ``Winston cone''
\cite{Win2005} designed to efficiently condense the light to the
output diameter of the light guide of 1$\ $cm matched to the active
area of a silicon photodiode. The use of a photodiode with a higher
quantum efficiency than a photomultiplier tube (afforded by larger
signals in the detection region) further improves the detection
efficiency. We estimate a detection efficiency of $\approx 40\%$ for
the atoms in the ${\rm 6s6p} \: ^3{\rm P}_0$, corresponding to
$\approx 26 \%$ of the atoms excited to the \TDOne.

Examples of signals obtained with the present apparatus are shown in
Figs. \ref{fig signal_no_B} and \ref{fig Signal_B}.

\section{Statistical sensitivity}
\label{Sec:Stat_Sens} Based on various parameters achieved in the
present apparatus, one can estimate the statistical sensitivity of
the PNC experiment. One straightforward way to do it is to note that
the Stark-interference technique is a tool for raising the signal
above the background and controlling systematic effects.  It does
not, however, improve the statistical sensitivity for a
shot-noise-limited experiment, but merely allows one to achieve the
shot-noise limit. Because of this, one can ignore the
Stark-interference nature of the experiment in the estimate of the
statistical sensitivity of the PNC measurement, and consider instead
the statistical sensitivity for direct excitation of the PNC
amplitude without any static electric field.

With $\approx 5 \: {\rm W}$ of circulating power in the
power-build-up cavity, using a detection efficiency of $26 \%$ as
discussed above, and a quantum efficiency of the photodiode of
$\approx 90\%$, we arrive at the number of detected transitions per
second of $\approx 0.7$. Thus, for a shot-noise-limited experiment,
the fractional precision is ${\delta E1_{\rm PNC} / E1_{\rm PNC}}
\approx {1 / \sqrt{\tau(s)}}$, where $\tau$ is the total integration
time of the experiment.  A realistic integration time of $10^{4} ~
{\rm s}$, gives a fractional precision of $\approx 1 \%$.

Achieving this statistical sensitivity seems quite realistic given
the current apparatus.  A preliminary analysis of systematic effects
indicates that systematic errors can be controlled at this level as
well.  Pushing the precision of the experiment to a level beyond
this point will require a significant amount of effort, most notably
a dramatic improvement in the power-build-up cavity in order to
achieve the necessary statistical sensitivity.  While it is
difficult to project the ultimate sensitivity of a PNC experiment in
Yb, the current status of the experiment gives us reason to hope
that a high precision experiment is possible.
\begin{figure}
\resizebox{0.45\textwidth}{!}{%
  \includegraphics{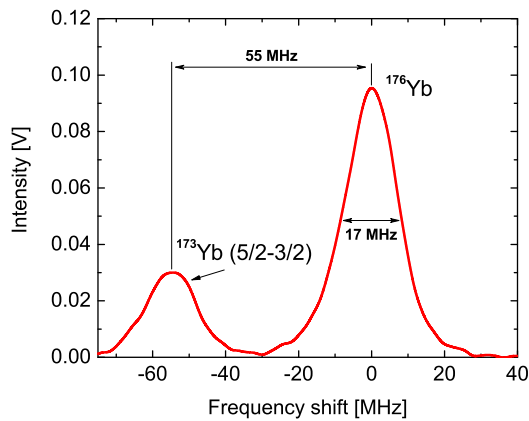}
} \caption{An example of the 556-nm fluorescence signal detected in
the interaction region recorded as the frequency of the 408-nm light
was swept across resonances corresponding to two neighboring
isotopic an hyperfine components of the Stark-induced transition.
E=12 kV/cm, B=0.} \label{fig signal_no_B}
\end{figure}
\begin{figure}
\resizebox{0.45\textwidth}{!}{%
  \includegraphics{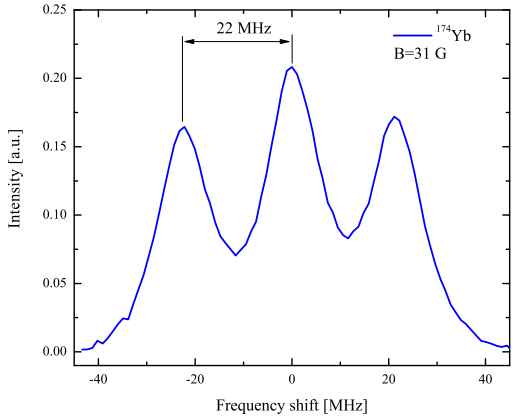}
} \caption{Same as in Fig. \ref{fig signal_no_B}, but the scan is
over the $^{174}$Yb resonance whose Zeeman components are split by
an applied magnetic field. The PNC signature is the relative
suppression/enhancement of the central peak with respect to the two
side peaks that is odd in E and even in B.} \label{fig Signal_B}
\end{figure}

\section{Analysis of systematics}
\label{Sec:Syst} The current Yb-PNC apparatus has been designed to
minimize systematic effects, and the majority of the systematic
effects we anticipate will result only in a combination of multiple
apparatus imperfections. Several reversals are available (electric
and magnetic fields, and light-polarization angle) to control
systematics. An additional helpful feature is a specific
characteristic light-detuning dependence of the PNC asymmetry.

Our general approach to dealing with systematics is similar to that
employed, for example, in Refs. \cite{Ngu97,Reg2002}. We begin by
making a list of all reasonable apparatus imperfections (field
misalignments, inhomogeneities, imperfect reversals, etc.), and
write out all possible P,T-even rotational invariants that produce
the same signature as the PNC effect upon reversals. We can then
artificially impose exaggerated combinations of imperfections in
order to measure (and, where possible, minimize) the remaining
imperfections. A detailed analysis of the systematics will be given
elsewhere. At present, we are confident that initially, systematics
should be under control at a $\sim 1\%$ level, sufficient for
reliable measurements of the isotopic and hyperfine PNC effects.

\section{Conclusions and acknowledgements}
\label{Sec:Conclusions}

We have reported on the progress in the experiment measuring
parity-nonconservation effects in ytterbium, which takes advantage
of an extraordinary anticipated enhancement of atomic PNC effects
(by two orders of magnitude), compared to, for example, Cs. The
Yb-PNC experiment relies on proven, although challenging,
experimental techniques of atomic physics. The atomic structure of
Yb is vastly different from the structure of atoms which have
previously been studied in atomic PNC experiments, so the specific
features of the experiment are quite different from the earlier
work, although there are many features drawn from earlier PNC
experiments \cite{Dre85,Woo99,Ngu97}. The seven stable isotopes of
Yb will allow for the first measurement of atomic PNC effects in a
chain of isotopes. Assuming the weak-charge variation between the
isotopes as predicted by the Standard Model, this will provide a
unique way of measuring the variation of neutron r.m.s. radii in an
isotopic chain. Conversely, if the neutron radii are known, one can
measure the isotopic variation of the nuclear weak charge
independent of uncertainties associated with atomic theory.
Measurements of the PNC effect on different hyperfine components of
the transition of atoms with non-zero nuclear spin nuclei
($^{171}$Yb and $^{173}$Yb) will allow for the first measurements of
the nuclear anapole moment in nuclei with unpaired neutrons.

The authors thank A.~Dilip, B.~P.~Das, and M.~G.~Kozlov for useful
discussions. This work has been supported by NSF (Grant 0457086),
and by the Director, Office of Science, Office of Basic Energy
Sciences, Nuclear Science Division, of the U.S. Department of Energy
under contract DE-AC03-76SF00098.

%
%
%

\bibliographystyle{epj}
\bibliography{JAtomicPNC}

\begin{thebibliography}{26}

\bibitem{Dem95}
D.~DeMille, Physical Review Letters \textbf{74}(21), 4165 (1995)

\bibitem{Por95}
S.G. Porsev, G.~Rakhlina~Yu, M.G. Kozlov, JETP Letters \textbf{61}(6), 459
  (1995)

\bibitem{Das97}
B.P. Das, Physical Review A \textbf{56}, 1635 (1997)

\bibitem{Woo99}
C.S. Wood, S.C. Bennett, J.L. Roberts, D.~Cho, C.E. Wieman, Canadian Journal of
  Physics \textbf{77}(1), 7 (1999)

\bibitem{Gue2005}
J.~Guena, M.~Lintz, M.A. Bouchiat, Physical Review A (General Physics)
  \textbf{71}, 042108 (2005)

\bibitem{Gin2004}
J.S.M. Ginges, V.~Flambaum, Physics Reports \textbf{397}(2), 63 (2004)

\bibitem{Hax2001}
W.C. Haxton, C.E. Wieman, in \emph{Annual Review of Nuclear and Particle
  Science} (2001), Vol.~51, pp. 261--293

\bibitem{Bow96}
C.J. Bowers, D.~Budker, E.D. Commins, D.~DeMille, S.J. Freedman, A.T. Nguyen,
  S.Q. Shang, M.~Zolotorev, Physical Review A \textbf{53}(5), 3103 (1996)

\bibitem{Bow99}
C.J. Bowers, D.~Budker, S.J. Freedman, G.~Gwinner, J.E. Stalnaker, D.~DeMille,
  Physical Review A \textbf{59}(5), 3513 (1999)

\bibitem{Sta2002}
J.E. Stalnaker, D.~Budker, D.P. DeMille, S.J. Freedman, V.V. Yashchuk, Physical
  Review A \textbf{66}(3), 31403 (2002)

\bibitem{Bud2003Jones}
D.~Budker, J.E. Stalnaker, Physical Review Letters \textbf{91}(26), 263901/1
  (2003)

\bibitem{Sta2006}
J.E. Stalnaker, D.~Budker, S.J. Freedman, J.S. Guzman, S.M. Rochester, V.V.
  Yashchuk, Physical Review A \textbf{73}, 043416 (2006)

\bibitem{Kim99}
D.F. Kimball, D.~Clyde, D.~Budker, D.~DeMille, S.J. Freedman, S.~Rochester,
  J.E. Stalnaker, M.~Zolotorev, Physical Review A \textbf{60}(2), 1103 (1999)

\bibitem{DeB2001}
B.~DeBoo, D.F. Kimball, C.H. Li, D.~Budker, Journal of the Optical Society of
  America B-Optical Physics \textbf{18}(5), 639 (2001)

\bibitem{Kim2001}
D.F. Kimball, Physical Review A \textbf{63}, 052113 (2001)

\bibitem{Bou75}
M.A. Bouchiat, C.~Bouchiat, Journal de Physique I \textbf{36}(6), 493 (1975)

\bibitem{Con79}
R.~Conti, P.~Bucksbaum, S.~Chu, E.D. Commins, L.~Hunter, Physical Review
  Letters \textbf{42}(6), 343 (1979)

\bibitem{Woo97}
C.S. Wood, S.C. Bennett, D.~Cho, B.P. Masterson, J.L. Roberts, C.E. Tanner,
  C.E. Wieman, Science \textbf{275}(5307), 1759 (1997)

\bibitem{Dre85}
P.S. Drell, E.D. Commins, Physical Review A \textbf{32}, 2196 (1985)

\bibitem{Sob92}
I.I. Sobelman, \emph{Atomic Spectra and Radiative Transitions}, Springer Series
  on Atoms and Plasmas, 2nd~edn. (Springer, New York, 1992)

\bibitem{Hoo2001}
C.J. Hood, H.J. Kimble, J.~Ye, Physical Review A \textbf{64}(3), 033804/1
  (2001)

\bibitem{Drev83}
R.W.P. Drever, J.L. Hall, F.V. Kowalski, J.~Hough, G.M. Ford, A.J. Munley,
  H.~Ward, Applied Physics B-Photophysics and Laser Chemistry \textbf{B31}(2),
  97 (1983)

\bibitem{Bud2000FP}
D.~Budker, S.M. Rochester, V.V. Yashchuk, Review of Scientific Instruments
  \textbf{71}(8), 2984 (2000)

\bibitem{Win2005}
R.~Winston, W.T. Welford, J.C. Miñano, P.~Benítez, \emph{Nonimaging optics}
  (Elsevier Academic Press, Amsterdam ; Boston, Mass., 2005)

\bibitem{Ngu97}
A.T. Nguyen, D.~Budker, D.~DeMille, M.~Zolotorev, Physical Review A
  \textbf{56}(5), 3453 (1997)

\bibitem{Reg2002}
B.C. Regan, E.D. Commins, C.J. Schmidt, D.~DeMille, Physical Review Letters
  \textbf{88}(7), 071805 (2002)

\end{thebibliography}
\end{document}